\documentclass[a4paper,11pt]{article}
\usepackage{pos}

\title{Scalar QED with Rydberg atoms}

\author*[a]{Yannick Meurice}
\author[a]{James Corona}
\author[b]{Sergio Cantu}
\author[b]{Fangli Liu}
\author[b]{Shengtao Wang}
\author[c]{Kenny  Heitritter}
\author[d]{Steve Mrenna}
\author[e]{Jin Zhang}
\author[f]{Shan-Wen Tsai}

\affiliation[a]{Department of Physics and Astronomy, University of Iowa, Iowa City, IA 52242, USA}
\affiliation[b]{QuEra Computing Inc., 1284 Soldiers Field Road, Boston, MA, 02135, USA}
\affiliation[c]{qBraid, 5235 S Harper Ct, Chicago, IL, 60615, USA}

\affiliation[d]{Fermi National Accelerator Laboratory,
P.O. Box 500, Batavia IL 60565, USA}
\affiliation[e]{Department of Physics and Chongqing Key Laboratory for Strongly Coupled Physics, Chongqing University, Chongqing 401331, China}
\affiliation[f]{Department of Physics and Astronomy, University of California, Riverside, CA 92521, USA}

\emailAdd{yannick-meurice@uiowa.edu}

\def\beq{\begin{equation*}}
\def\enq{\end{equation*}}

\abstract{

We review recent suggestions to quantum simulate scalar electrodynamics (the lattice Abelian Higgs
model) in $1+1$ dimensions with rectangular arrays of Rydberg atoms. We show that platforms made
publicly available recently allow empirical explorations of the critical behavior of quantum simulators. We discuss recent progress regarding the phase diagram of two-leg ladders, effective Hamiltonian approaches and the construction of hybrid quantum algorithms targeting hadronization in
collider physics event generators.
\vskip10pt
FERMILAB-CONF-23-0831-CSAID

 }

\FullConference{The 40th International Symposium on Lattice Field Theory (Lattice 2023)\\
July 31st - August 4th, 2023\\
Fermi National Accelerator Laboratory\\}

\begin{document}
\newcommand{\T}{\text{T}}
\newcommand{\bra}[1]{\ensuremath{\langle #1 |}}
\newcommand{\ket}[1]{\ensuremath{| #1 \rangle}}
\newcommand{\braket}[2]{\ensuremath{\langle #1 | #2 \rangle}}
\renewcommand{\dag}{\dagger}
\maketitle

\section{Introduction}
In these Proceedings we discuss the possibility of hybrid quantum/classical computing for event generators such as PYTHIA
\cite{anderssonPartonFragmentationString1983,bierlichComprehensiveGuidePhysics2022}. Our long-term goal is to replace the current hadronization part which turns quarks and gluons into hadrons and is currently implemented with the phenomenological Lund model, by an ab-initio lattice calculation performed by a quantum computing device. More technical details and references  can be found in \cite{Heitritter:2022jik}. 

Recently there has been a lot of interest for quantum simulation for gauge theories 
\cite{Dalmonte:2016alw,Banuls:2019bmf,Kasper:2020akk,Bauer:2022hpo,Halimeh:2023lid,DiMeglio:2023nsa} and in particular for the real-time dynamics. In the following, we consider the possibility where the hadronization part of event generators could be  described by a simplified lattice model: the compact Abelian Higgs Model (Scalar QED) in $1+1$ dimensions. We explain that is (literally!) possible to attempt model building for a quantum simulator with arrays of Rydberg atoms \cite{51qubits,keesling2019,Browaeys:2020kzz}: publicly available interfaces \cite{quera} allow users to engage in quantum simulations in a rather straightforward way. 
The matching between the target model and the simulator is non-trivial. However, we show that for some region of the parameters of the simulator, it is possible to construct an effective theory for the simulator which has only one (important) term differing from the target model. More details about this question can be found in a recent preprint \cite{Zhang:2023agx}. We also briefly discuss practical examples of current  work with QuEra.

\section{The compact Abelian Higgs Model (CAHM)}
\begin{figure}
\begin{center}
\includegraphics[width=3.5in]{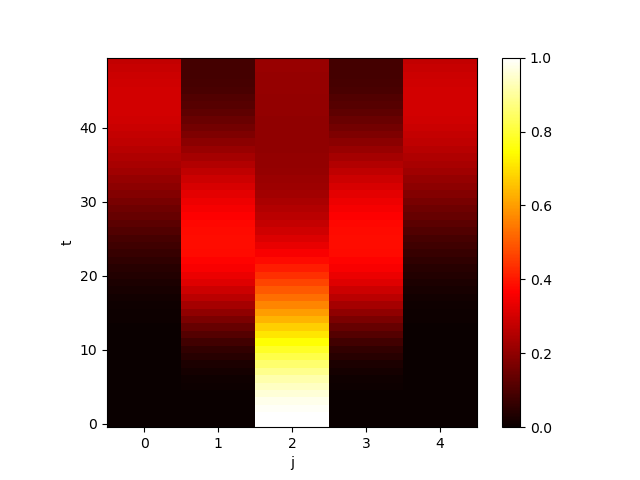}
\end{center}
\caption{Electric field $E_j$ versus position $j$ for the evolution of a particle-antiparticle pair for the Abelian Higgs model in $1+1$ dimensions for $Y/X=1$ and $U/X=0.1$ and 
50 time steps with $\Delta t$ such that $X\Delta t=0.1$.}
\label{fig1}
\end{figure}

The lattice compact Abelian Higgs model is a non-perturbative regularized formulation of scalar quantum electrodynamics 
(scalar electrons-positrons + photons with compact fields). The partition function reads: 
\beq
Z_{CAHM}=\prod_x\int_{-\pi}^{\pi}\frac{d\varphi _x}{2\pi}\prod_{x,\mu}\int_{-\pi}^{\pi}\frac{dA_{x,\mu}}{2\pi}
e^{-S_{gauge}-S_{matter}}, 
\label{eq:gaugemeasure}
\enq
with
\beq
S_{gauge}=\beta_{plaquette}\sum_{x,\mu<\nu} (1-\cos(A_{x,\mu}+A_{x+\hat{\mu},\nu}-A_{x+\hat{\nu}, \mu}-A_{x,\nu})),
\label{eq:gauge}
\enq
\beq
\label{eq:smatter}
S_{matter}=\beta_{link} \sum\limits_{x,\mu} (1-\cos(\varphi_{x+\hat{\mu}}-\varphi_x+A_{x,\mu})).\enq
It has a local invariance: 
$\varphi_x'=\varphi_x+\alpha_x$ and 
where local changes in $S_{matter}$ are compensated by
$A_{x,\mu}'=A_{x,\mu}-(\alpha_{x+\hat{\mu}}-\alpha_x)$.
$\varphi$ is the Nambu-Goldstone mode of the original model. 
The Brout-Englert-Higgs mode is decoupled (heavy). 
Using methods of Tensor Lattice Field Theory (TLFT) reviewed in \cite{meurice2020}, one obtains an 
Hamiltonian and Hilbert space in $1+1$ dimension
in the continuous-time limit:
$$H = \frac{U}{2}\sum_{i=1}^{N_{s}} \left(L^z_{i}\right)^2 
	+ \frac{Y}{2} {\sum_i}  (L^z_{i+1} - L^z_{i})^2-
	X\sum_{i=1}^{N_{s}} U^x_{i} \ $$
with 
$U^x\equiv \frac{1}{2}(U^+ + U^-) $ and 
$L^z\ket{m}=m\ket{m} \ \ {\rm and}\ U^\pm\ket{m}=\ket{m\pm1}.$
$m$ is a discrete electric field quantum number ($-\infty< m<+\infty$). In practice, we need to apply  truncations:  
For a spin-$m_{max}$ truncation we have 
$U^\pm\ket{\pm m_{max}}=0.$
We focus on the spin-1 truncation ($m = \pm 1, 0$ and  $U^x=L^x/\sqrt{2}$).
The $U$-term represents the electric field energy, the 
$Y$-term, the matter charges (determined by Gauss's
law) and the $X$-term: currents inducing temporal changes in the electric field. Using this Hamiltonian, we have evolved an intial state with a bit of electric field in the middle. By Gauss's law, this is equivalent to a pair of bosonic particle-antiparticle separated by one lattice spacing. An example is shown in Fig.~\ref{fig1}. and various stages can be reinterpreted in terms of patterns such as string breaking, discussed in Refs. \cite{anderssonPartonFragmentationString1983,bierlichComprehensiveGuidePhysics2022}. 

\section{The simulator}

\begin{figure}
\begin{center}
\includegraphics[width=3.5in]{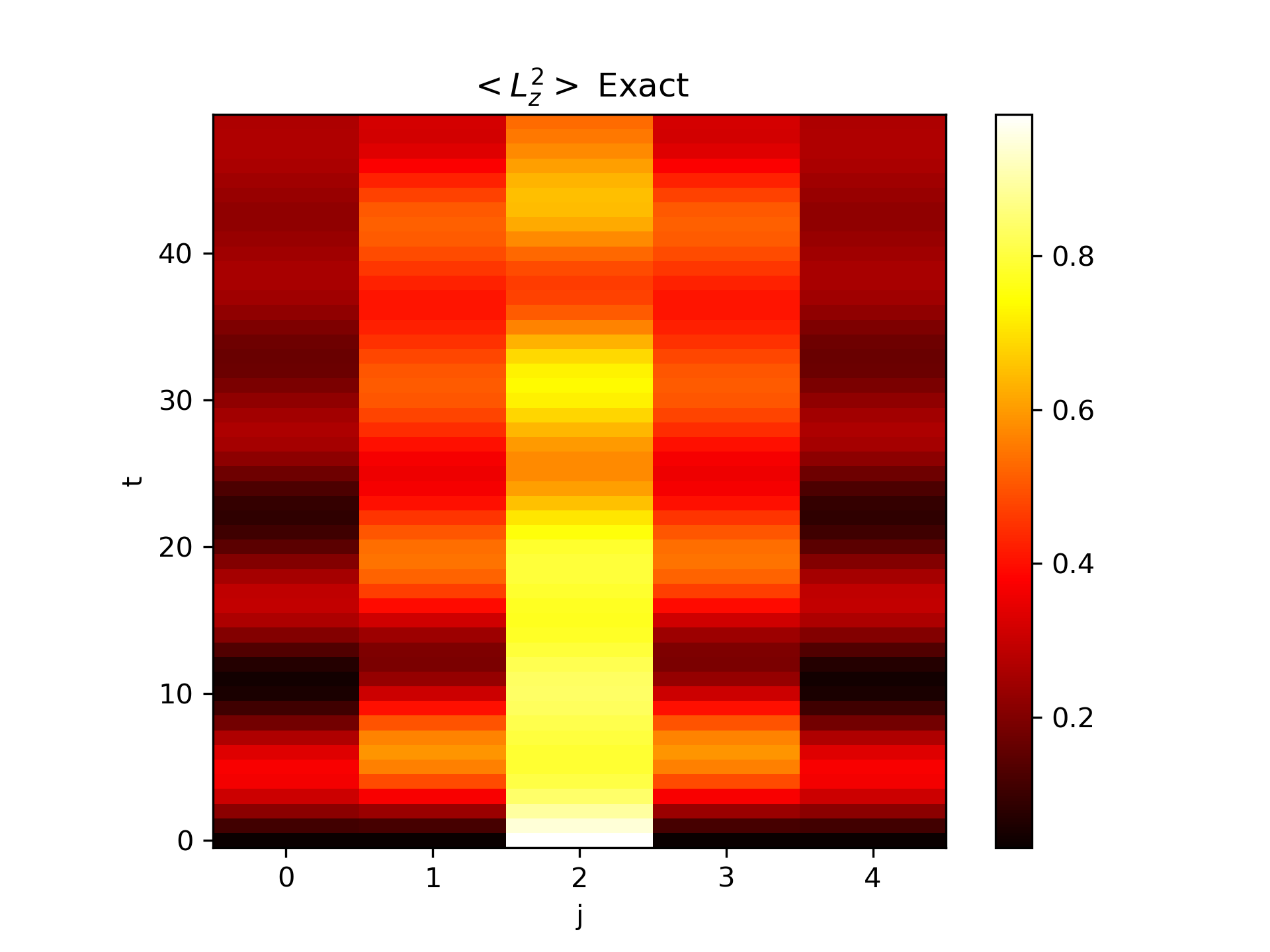}
\caption{Time evolution for $E^2_j=(n_j^{up}-n_j^{down})^2$ for a 5-rung ladder (ten $^{87}Rb$ atoms). At initial time, all the atoms are in the ground state except for the top atom in the middle rung.
$\Omega=2\pi \times 2*10^6$ MHz, 
$\Delta=2\Omega$, $a_x=R_b$ $a_y=0.5 R_b$.}
\label{fig2}
\end{center}
\end{figure}

In \cite{cara}, we adapted the optical lattice construction \cite{Zhang:2023agx} 
using arrays of $^{87}Rb$ atoms \cite{51qubits,keesling2019,Browaeys:2020kzz,quera} separated by controllable (but not too small) distances, coupled to the excited Rydberg state $\ket{r}$ with a detuning $\Delta$. 
The ground state is denoted $\ket{g}$ 
and the two possible states  $\ket{g}$ and $\ket{r}$ can be seen as a qubit. We have the occupations $n\ket{g}=0, \ n\ket{r}=\ket{r}$. 
The Hamiltonian reads 
\beq
\label{eq:genryd}
H = \frac{\Omega}{2}\sum_i(\ket{g_i}\bra{r_i} + \ket{r_i}\bra{g_i})-\Delta\sum_i  n_i +\sum_{i<j}V_{ij}n_in_j,
\enq
with 
\beq\label{eq:oneosix}
V_{ij}=\Omega R_b^6/r_{ij}^6,\enq
for a distance $r_{ij}$ between the atoms labelled as $i$ and $j$. Note that when $r=R_b$, the Rydberg radius, $V=\Omega$. 
This repulsive interaction prevents two atoms close enough to each other to be both in the $\ket{r}$ state. This is the
so-called blockade mechanism which can be used to produce an effective spin-1 local Hilbert space. In order to shortcut the discussion of the staggered interpretation of the electric field, we have plotted its square $E^2_j=(n_j^{up}-n_j^{down})^2$ (see \cite{cara} for a discussion) in Fig.~\ref{fig2} which also exhibit interesting behavior with a similar interpretation as for Fig.~\ref{fig1}. 

\section{The effective Hamiltonian and phase diagram}

As reported in a recent preprint \cite{Zhang:2023agx}, we constructed a translation-invariant effective Hamiltonian by integrating over the simulator high-energy states produced by the blockade mechanism. Remarkably, for all the simulators considered (ladders, prisms, and others), the effective Hamiltonians have the three types of terms present for the CAHM (Electric field, matter charge and currents energies) but, in addition, terms quartic in the electric field. For positive detuning, the new terms create degenerate vacua resulting in a very interesting phase diagram. 

The phase diagram of the ladder simulator with rung size twice the lattice spacing has been investigated experimentally. 
Evidence for an incommensurate phase between crystalline commensurate phases will be reported in an upcoming preprint. 
It is relatively straightforward for theoretical users to perform analog simulations with Rydberg arrays using 
publicly available interfaces \cite{quera}. We hope that this exploration will improve our understanding of 
inhomogeneous phases and the Lifshitz regime of lattice quantum chromodynamics \cite{Pisarski:2019cvo, Kojo:2009ha}, as well as the examination of ``chiral spiral” condensation~\cite{Basar:2008im}. One should also compare the manipulation of the three states associated with a rung in our approach with qutrit simulations \cite{Gustafson:2021qbt} and decide  if the quartic term could be found in the context of Symanzik improvement \cite{Carena:2022kpg,Carena:2022hpz}.

\section{Conclusions}

We have considered ladder-shaped Rydberg arrays with two atom per rung as simulators  for the compact Abelian Higgs model.
Ultimately the matching between simulator and target model should be understood in the continuum limit (universal behavior).
Effective Hamiltonians for the simulator were found with same three types of terms as the 
target model plus an extra quartic term.
The two-leg ladder has a very rich phase diagram.
Explorations of the phase diagram with AWS/QuEra are ongoing as well as the possibility of an hybrid interface with PYTHIA are pursued. 

\section*{Acknowledgemnts}
Y.M. and K.H. were supported in part by the Dept.~of Energy under Award Number DE-SC0019139. Y. M. thanks the Amazon Web Services and S. Hassinger for facilitating remote access to QuEra through the Amazon Braket.
K.H. acknowledges support from the URA Visiting Scholars Program. J.Z. is supported by NSFC under Grants No.~12304172 and No.~12347101, Chongqing Natural Science Foundation under Grant No.~CSTB2023NSCQ-MSX0048, and Fundamental Research Funds for the Central Universities under Projects No.~2023CDJXY-048 and No.~2020CDJQY-Z003. 
This work was also supported in part by the National Science Foundation (NSF) RAISE-TAQS under Award Number 1839153 (SWT). Computations were performed using the computer clusters and data storage resources of the HPCC, which were funded by grants from NSF (MRI-1429826) and NIH (1S10OD016290-01A1).
S.M. is supported in part by the U.S. Department of Energy, Office of Science, Office of High Energy Physics QuantISED program under the grants ``HEP Machine Learning and Optimization Go Quantum'', Award Number 0000240323, and ``DOE QuantiSED Consortium QCCFP-QMLQCF'', Award Number DE-SC0019219. This manuscript has been authored by Fermi Research Alliance, LLC under Contract No.~DEAC02-07CH11359 with the U.S. Department of Energy, Office of Science, Office of High Energy Physics. This research was supported in part through computational resources provided by The University of Iowa, Iowa City, Iowa.

\end{document}